\documentclass[conference]{IEEEtran}
\usepackage{amsmath, mathrsfs,balance}
\usepackage{bm}

\usepackage{graphicx}
\usepackage{color}
\usepackage{tikz}
\usepackage{pgfplots}
\pgfplotsset{compat=newest}
\usetikzlibrary{patterns}
\usetikzlibrary{positioning}
\usetikzlibrary{datavisualization}
\usetikzlibrary{datavisualization.formats.functions}
\usetikzlibrary{backgrounds}
\usetikzlibrary{shapes,snakes}

\makeatletter
\def\maketag@@@#1{\hbox{\m@th\normalfont\normalsize#1}}
\makeatother

\usepackage[none]{hyphenat}

\newcommand{\subparagraph}{}
\usepackage[compact]{titlesec}
\titlespacing*{\section}{2pt}{1\baselineskip}{0.9\baselineskip}

\allowdisplaybreaks

\usepackage{amsfonts}
\usepackage{amssymb}
\usepackage{color}
\usepackage{multirow}
\usepackage{graphicx}

\usepackage{caption}
\usepackage{subcaption}
\usepackage[numbers,sort&compress]{natbib}
\usepackage{balance}

\usepackage{mathbbol}

\def\mindex#1{\index{#1}}

\def\sq{\hbox{\rlap{$\sqcap$}$\sqcup$}}
\def\qed{\ifmmode\sq\else{\unskip\nobreak\hfil
\penalty50\hskip1em\null\nobreak\hfil\sq
\parfillskip=0pt\finalhyphendemerits=0\endgraf}\fi\medskip}

\long\def\defbox#1{\framebox[.9\hsize][c]{\parbox{.85\hsize}{%
\parindent=0pt
\baselineskip=12pt plus .1pt      
\parskip=6pt plus 1.5pt minus 1pt 
 #1}}}

\long\def\beginbox#1\endbox{\subsection*{}%
\hbox{\hspace{.05\hsize}\defbox{\medskip#1\bigskip}}%
\subsection*{}}

\def\endbox{}

\newsavebox{\junk}
\savebox{\junk}[1.6mm]{\hbox{$|\!|\!|$}}

\def\bC{{\mathbb C}}

\def\sfC{{\sf C}}

\def\bfmath#1{{\mathchoice{\mbox{\boldmath$#1$}}%
{\mbox{\boldmath$#1$}}%
{\mbox{\boldmath$\scriptstyle#1$}}%
{\mbox{\boldmath$\scriptscriptstyle#1$}}}}

\def\bfmY{\bfmath{Y}}

\def\bfmhhaY{\bfmath{\hhaY}} 
\def\bfmhhaY{\hbox to 0pt{$\widehat{\bfmY}$\hss}\widehat{\phantom{\raise 1.25pt\hbox{$\bfmY$}}}}

\def\til={{\widetilde =}}

\def\clA{{\cal A}}

\def\clC{{\cal C}}
\def\clD{{\cal D}}

\def\clG{{\cal G}}

\def\clL{{\cal L}}

\def\clN{{\cal N}}

\def\clR{{\cal R}}

\def\clT{{\cal T}}

 \def\FRAC#1#2#3{\genfrac{}{}{}{#1}{#2}{#3}}

\def\fraction#1#2{{\mathchoice{\FRAC{0}{#1}{#2}}%
{\FRAC{1}{#1}{#2}}%
{\FRAC{3}{#1}{#2}}%
{\FRAC{3}{#1}{#2}}}}

\def\ddtp{{\mathchoice{\FRAC{1}{d^{\hbox to 2pt{\rm\tiny +\hss}}}{dt}}%
{\FRAC{1}{d^{\hbox to 2pt{\rm\tiny +\hss}}}{dt}}%
{\FRAC{3}{d^{\hbox to 2pt{\rm\tiny +\hss}}}{dt}}%
{\FRAC{3}{d^{\hbox to 2pt{\rm\tiny +\hss}}}{dt}}}}

\def\average#1,#2,{{1\over #2} \sum_{#1}^{#2}}

\def\eye(#1){{\bf(#1)}\quad}

\def\eq#1/{(\ref{e:#1})}

\newcommand{\beqn}[1]{\notes{#1}%
\begin{eqnarray} \elabel{#1}}

\newcommand{\eeqn}{\end{eqnarray} }

\newcommand{\beq}[1]{\notes{#1}%
\begin{equation}\elabel{#1}}

\newcommand{\eeq}{\end{equation}}

\def\bdes{\begin{description}}
\def\edes{\end{description}}

\newcounter{rmnum}

\newcounter{anum}

{\end{list}}

\def\ass(#1:#2){(#1\ref{#1:#2})}

\def\ritem#1{
\item[{\sf \ass(\current_model:#1)}]
}

\newenvironment{recall-ass}[1]{%
\begin{description}
\def\current_model{#1}}{
\end{description}
}

\long\def\comment#1{}

\renewcommand{\arg}{{\hbox{arg}}}

\usepackage{tikz}
\usepackage{pgfplots}
\pgfplotsset{compat=newest}
\usetikzlibrary{patterns}
\usetikzlibrary{positioning}
\usetikzlibrary{datavisualization}
\usetikzlibrary{datavisualization.formats.functions}
\usetikzlibrary{backgrounds}
\usetikzlibrary{shapes,snakes}
\usepackage{textcomp}
\usetikzlibrary{automata}
\usepgfplotslibrary{groupplots}

\allowdisplaybreaks

\usepackage[ruled]{algorithm2e}
\usepackage{algpseudocode}
\usepackage{pifont}
\usepackage{varwidth}

\usepackage{multicol,lipsum}
\usepackage{array}

\usepackage{lscape}

\def\cg{{\clC\clN}}

\usepackage{color}
\newcommand{\figref}[1]{Fig.~\ref{#1}}
\usetikzlibrary{decorations.markings}

\IEEEoverridecommandlockouts

\begin{document}

\sloppy

\title{Queue-Aware Beam Scheduling for Half-Duplex mmWave Relay Networks}
\author{\IEEEauthorblockN{Xiaoshen Song, Giuseppe Caire}
	
\IEEEauthorblockA{Communications and Information Theory Chair, Technische Universit\"at Berlin, Germany}
}

\maketitle

\begin{abstract}
Millimeter wave (mmWave) bands are considered a powerful key enabler for next generation (5G) mobile networks by providing multi-Gbps data rates. However, their severe pathloss and sensitivity to blockage present challenges in practical implementation. One effective way to mitigate these effects and to increase communication range is beamforming in combination with relaying. In this paper, we focus on two typical mmWave relay networks and for each network, we propose three beam scheduling methods to approach the network information theoretic capacity. The proposed beam scheduling methods include the deterministic horizontal continuous edge coloring (HC-EC) scheduler, the adaptive back pressure (BP) scheduler and the adaptive low-delay new back pressure (newBP) scheduler. With the aid of computer simulations, we show that within the network capacity range, the proposed schedulers provide good guarantees for the network stability, meanwhile achieve very low packet end-to-end delay.
\end{abstract}	

\begin{IEEEkeywords}
mmWave, relay network, network stability, end-to-end delay, network capacity 
\end{IEEEkeywords}

\section{Introduction}\label{introduction}
With the explosive growth of mobile data demand, the 5th generation (5G) mobile networks with carrier frequency in the millimeter wave (mmWave) bands ($30$-$300\,$GHz) represent an attractive complement to networks operating in conventional cellular frequency bands \cite{Xingqin5G2019}. Due to the large available bandwidth, a mmWave transceiver can potentially achieve individual link rates in tens of Gbps. However, mmWave transmissions suffer from high propagation loss and penetration loss, resulting in very limited range/coverage and very high susceptibility to blockage. One effective way to mitigate these effects and to increase the communication range is beamforming in combination with relaying \cite{Petropulu2019relay}, where the former is achieved by utilizing large antenna arrays on the transceivers and pointing their beams towards each other, and the latter refers to using intermediate nodes to relay the source signal to the destination \cite{Yan2018}.

The beamforming problem in a small cell mmWave scenario with one base station and multiple users has been studied in our previous work \cite{sxsBA2017, sxs2018TimeJour,sxs2019OSPS}. With the increasing interest in developing small cells for mmWave communication, how to use relays to increase the coverage to support mmWave wireless backhaul for dense small cell deployments is a major challenge \cite{Yan2018}.

In a relay-assisted mmWave network, the relay nodes divide the long link into some short but very high rate links to overcome the mmWave sensitivity to blockage. In such a situation, a link is active only if both nodes focus their beams to face each other, which is determined by the underlying beam scheduling scheme. The source and destination cannot communicate with each other directly because the distance between them is too large to achieve the required data rate and/or some obstacles are in between preventing direct communication. Given a set of static relays between a source-destination pair, several relay selection schemes for two-hop or multi-hop mmWave settings have been proposed in recent literature \cite{Petropulu2019relay,Yan2018,Hamdi2016FD,Omer2019FL}. However, these works are limited in the selection of a single relay path with the largest end-to-end signal to noise ratio (SNR), the utilization of multiple relay paths in combination with an efficient beam scheduling scheme to further increase the network throughput is yet unexplored. 

\begin{figure}[t]
	\centering
	\scalebox{0.75}{\usetikzlibrary{arrows.meta}

\begin{tikzpicture}
\draw[help lines,color=white,opacity=0] (0,0) grid (8,7);

\begin{scope}[transform canvas={xshift = 0cm, yshift = 5cm}]
\draw[opacity=0,fill=red,fill opacity=0.3] (0.75, 1) ellipse (0.75cm and 0.25cm);
\draw[opacity=0,fill=green,fill opacity=0.5] (1.75, 1) ellipse (0.75cm and 0.25cm);

\draw[opacity=0,fill=red,fill opacity=0.3] (4.75, 1) ellipse (0.75cm and 0.25cm);
\draw[opacity=0,fill=green,fill opacity=0.5] (5.75, 1) ellipse (0.75cm and 0.25cm);

\draw[thick,fill=white] (0,1) circle (0.2);
\draw[thick,fill=white] (2.5,1) circle (0.2);
\draw[thick,fill=white] (4,1) circle (0.2);
\draw[thick,fill=white] (6.5,1) circle (0.2);
\draw[thick,fill=white] (8,1) circle (0.2);

\draw[-Latex, thick, dashed, color=red] (0.2,1)--(2.3,1);
\draw[-Latex, thick, dashed] (2.7,1)--(3.8,1);
\draw[-Latex, thick, dashed,color=red] (4.2,1)--(6.3,1);
\draw[-Latex, thick, dashed] (6.7,1)--(7.8,1);

\node[above] at (0,1.2) {\large S};
\node[above] at (8,1.2) {\large D};
\node[above] at (4,0) {\large (a)};
\end{scope}

\begin{scope}[transform canvas={xshift = 0cm, yshift = 1cm}]

\draw[-Latex, thick, dashed,color=red] (1,1.5)--(3.8,3);
\draw[-Latex, thick, dashed] (1,1.5)--(3.8,2);
\draw[-Latex, thick, dashed] (1,1.5)--(3.8,1);
\draw[-Latex, thick, dashed] (1,1.5)--(3.8,0);

\draw[-Latex, thick, dashed] (4.2,3)--(6.8,1.8);
\draw[-Latex, thick, dashed] (4.2,2)--(6.8,1.6);
\draw[-Latex, thick, dashed,color=red] (4.2,1)--(6.8,1.4);
\draw[-Latex, thick, dashed] (4.2,0)--(6.8,1.2);

\draw[opacity=0,fill=red,fill opacity=0.3,rotate around={28:(1, 1.5)}] (1.95, 1.5) ellipse (0.95cm and 0.3cm);
\draw[opacity=0,fill=green,fill opacity=0.5,rotate around={25:(4, 3)}] (3.05, 3) ellipse (0.95cm and 0.3cm);

\draw[opacity=0,fill=red,fill opacity=0.3,rotate around={8.6:(4, 1)}] (4.9, 1) ellipse (0.9cm and 0.25cm);
\draw[opacity=0,fill=green,fill opacity=0.5,rotate around={10:(7, 1.5)}] (6.1, 1.5) ellipse (0.9cm and 0.25cm);

\draw[thick,fill=white] (1,1.5) circle (0.2);
\draw[thick,fill=white] (7,1.5) circle (0.2);
\draw[thick,fill=white] (4,0) circle (0.2);
\draw[thick,fill=white] (4,1) circle (0.2);
\draw[thick,fill=white] (4,2) circle (0.2);
\draw[thick,fill=white] (4,3) circle (0.2);

\node[left] at (0.7,1.5) {\large S};
\node[right] at (7.3,1.5) {\large D};
\node[above] at (4,-1) {\large (b)};

\end{scope}
\end{tikzpicture}}
	\caption{\small An illustration of the underlying mmWave relay networks: (a) The line network $\clL$, with two links being activated (the red dashed link covered by a beam pair); (b) The diamond network $\clD$, with two links being activated (the red dashed link covered by a beam pair).}
	\label{network_model}
\end{figure}
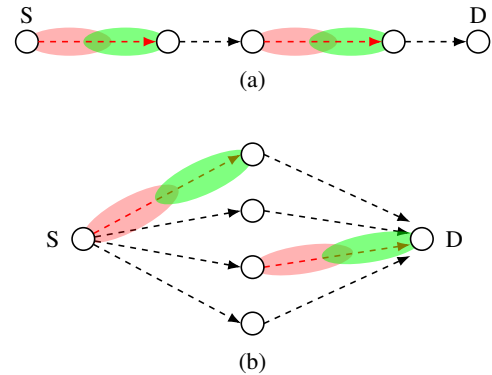
A recent study of 1-2-1 model in \cite{Yahya2017Line, Yahya2018Diam} provides an idealized and simplified information theoretic relay network model, where a potential link is active only if the transmitter beam and the receiver beam are pointing at each other which is termed as 1-2-1 link in \cite{Yahya2018Diam}. By noticing the great similarities between the 1-2-1 link and the mmWave communication link (i.e., very high pathloss and directivity), in this paper, we focus on two basic mmWave relay networks, i.e., the line network and the diamond network as shown in \figref{network_model} (a) and (b), respectively. In both cases, the relay nodes are assumed to work in half duplex (HD) mode that cannot simultaneously transmit and receive. A link is active only if both nodes focus their beams to face each other. We go beyond the limitations in most existing works \cite{Petropulu2019relay,Yan2018,Hamdi2016FD,Omer2019FL} which have considered only single path selection. Here we take the general optimality of the beam scheduling approach in \cite{Yahya2017Line, Yahya2018Diam} and find actual computational algorithms (both offline and online) to approach the beam scheduling theory in \cite{Yahya2017Line, Yahya2018Diam}, which is information theoretically optimal for the 1-2-1 networks within a gap that depends only on the network size but not on the link capacities (i.e., on SNR).
We propose three beam scheduling methods: the deterministic horizontal continuous edge coloring (HC-EC) scheduler, the adaptive back pressure (BP) scheduler and the adaptive low-delay new back pressure (newBP) scheduler. We provide a joint performance evaluation of the network stability and the end-to-end delay in terms of different source input rates. It is shown through simulations that all the three beam scheduling methods are throughput optimal$\,$\footnote{A scheduling method is throughput-optimal if it makes all the network queues stable (i.e., bounded in terms of average size) for all source arrival rates $\bar{A}_0\!<\!\sfC$ \cite{Tassiulas1990Stability}, where $\sfC$ is the information theoretic relay network capacity.} and achieve very low end-to-end delay. Particularly, each of the proposed beam scheduling methods has its own superiority in terms of different source input rates and computation complexity \cite{Yahya2017Line, Yahya2018Diam}.    

The reminder of the paper is organized as follows. In Section \ref{systemmodel}, we introduce the network model and the problem formulation. Section \ref{SchedulingMethod} presents our proposed beam scheduling methods. Section \ref{simulations} provides numerical results and Section \ref{conclusion} concludes this paper.

\section{System Model}\label{systemmodel}
In this section, we present the formulation of two 1-2-1 network topologies considered in this paper. These topologies can be seen as basic building blocks for more general topologies.
\subsection{The line network $\clL$}
For the line network, we consider the $N$-relay Gaussian HD channel model as shown in \figref{network_model}$\,$(a), where a source node (node $0$) wishes to communicate to a destination node (node $N+1$) through a route of $N$ relays. Each relay is operating in HD. Denoted by $h_{i,j}$ as the complex channel coefficients from node $j$ to node $i$ and $h_{i,j}=0$ whenever $j\neq i-1$. The channel gains are assumed to be constant for the whole transmission duration. We assume that the channel inputs satisfy an unit average power constraint, hence the point-to-point link capacity from node $i-1$ to node $i$ can be written as
\begin{align}
l_i=\log (1+|h_{i,i-1}|^2), \quad \forall i\in [N+1],
\end{align}  
where we assume the additive white Gaussian noise at each nodes are independent and identically distributed (i.i.d.) as $\cg(0, 1)$, and where $[N]$ indicates the set of non-negative integers $\{1,..., N\}$. Following \cite{Yahya2017Line}, the capacity of the Gaussian HD line network can be described to within a constant gap $\text{GAP}_{\clL}=O(N)$ as
\begin{align}\label{cap_line}
\sfC_{\clL} = \max_{\lambda\in\Lambda} \min_{\clA\subseteq[N]} \sum_{s\in\{0,1\}^{N}}\lambda_{s}\sum_{\substack{i\in\{N+1\}\cup\{\clR_{s}\cap\clA\}\\i-1\in\{0\}\cup\{\clT_{s}\cap\clA^{c}\}}} l_{i}.
\end{align}
Such constant gap result is referred to as the network {\em approximate capacity} in \cite{Yahya2018Diam}. Here the schedule $\lambda\in\bC^{2^N}$ determines the fraction of time $\lambda_{s}$ for each network state $s\in\{0,1\}^{N}$, $\Lambda=\{\lambda:\lambda\in\bC^{2^N},\lambda\geq 0,\|\lambda\|_{1}=1\}$ is the set of all possible schedules, $\clR_{s}$ (w.r.t. $\clT_{s}$) represents the set of indices of receiving (w.r.t. transmitting) relays in the state $s$, and $\clA^{c}=[N]\setminus \clA$. In \cite{Yahya2017Line}, the authors propose a simple edge coloring (EC) scheduler that achieves the {\em approximate capacity}. It follows that the {\em approximate capacity} in \eqref{cap_line} is given explicitly by \cite{Yahya2017Line}
\begin{align}\label{cap_line_simp}
\sfC_{\clL} = \min_{i\in[N]} \left\{\fraction{l_{i}\cdot l_{i+1}}{l_{i}+l_{i+1}}\right\}.
\end{align} 
The EC scheduler presented in \cite{Yahya2017Line} leverages the similarities between network states in HD and edge coloring in a graph. More precisely, an edge coloring assigns colors to edges in a graph such that no two adjacent edges are colored with the same color. Similarly in HD, a network node cannot be a receiver and a transmitter simultaneously. It follows that the EC scheduler assigns one color to potential links to indicate that these links are activated in the underlying network state. Although the EC scheduler in \cite{Yahya2017Line} achieves the {\em approximate capacity} $\sfC_{\clL}$ \eqref{cap_line_simp}, it ignores the average queues over each network node, which may result in a very large transmission delay. In Section \ref{SchedulingMethod}, we will provide a modified version of the EC scheduler, which can significantly reduce the transmission delay.

\subsection{The diamond network $\clD$}
Similar with the line network $\clL$, for the diamond network $\clD$, we consider the $N$-relay Gaussian HD channel model as shown in \figref{network_model}$\,$(b), where $N$ relays assist the communication between a source node (node $0$) and a destination node (node $N+1$). More precisely, the underlying diamond consists of $N$ paths with in total $2N$ links. Each path consists of the source node, one relay and the destination node. Due to the HD constraint, the two links w.r.t. one relay cannot be activated simultaneously. Denoted by $h_{p,1}$ and $h_{p,2}$, $p\in[N]$, as the complex channel coefficients in the first and the second hops, respectively. Consider the unite input power constraints and the i.i.d. Gaussian noise of $\cg(0, 1)$ at each nodes, the point-to-point link capacity in the diamond network can be written as
\begin{align}
l_{p,j}=\log (1+|h_{p,j}|^2), \quad \forall p\in [N], \quad \forall j\in [2].
\end{align} 
It has been shown in \cite{Yahya2018Diam} that the {\em approximate capacity} of the $N$-relay HD Gaussian diamond network $\clD$ can be written as
\begin{align}\label{cap_diam}
\sfC_{\clD} &= \max \sum_{p\in[N]}x_{p}\sfC_{p}\nonumber\\
&s.t.\quad  0\leq x_{p} \leq 1, \quad \forall p\in [N]\nonumber\\
&\quad\quad \sum_{p\in[N]}x_{p}\frac{\sfC_{p}}{l_{p,1}}\leq 1,\nonumber\\
&\quad\quad \sum_{p\in[N]}x_{p}\frac{\sfC_{p}}{l_{p,2}}\leq 1,
\end{align}
where $x_{p}$ represents the fraction of time that the $p$-th path ( $0\rightarrow p \rightarrow N+1$) is utilized in the network, and $\sfC_{p}$ is the capacity of $p$-th path, given by 
\begin{align}\label{path_cap_diam}
\sfC_{p}=\frac{l_{p,1}\cdot l_{p,2}}{l_{p,1}+ l_{p,2}}.
\end{align}
It has been proved in \cite{Yahya2018Diam} that for the $N$-relay HD Gaussian diamond network $\clD$, the {\em approximate capacity} $\sfC_{\clD}$ can always be achieved by activating at most $3$ relays in the network, independently of $N$.

\subsection{Network stability and end-to-end delay}
We say that a network is stable for a source arrival rate $\bar{A}_0$ if there exists a transmission strategy such that the average backlog of all queues is finite. A well known result \cite{Neely2006backpressure} is that the network could be stable for any $\bar{A}_0\!<\!\sfC_{\clL}$ (w.r.t. $\bar{A}_0\!<\!\sfC_{\clD}$). 
In this paper, we assume that the network operates in slotted time, denoted by $t\geq 0$. Considering a first-in-first-out (FIFO) system, we assume that only the packets currently in node $i$ at the beginning of slot $t$ can be transmitted during that slot. Denoted by $A_i(t)$ and $D_i(t)$ as the number of arrival packets and the number of departure packets at node $i$, respectively. The slot-to-slot dynamics of the queuing backlog $U_{i}{(t)}$ satisfies the following:
\begin{align}\label{queue_dynamic}
U_{i}{(t+1)} = \max\big\{U_{i}{(t)}-D_i(t), 0\big\} + A_i(t)
\end{align}
To evaluate the network stability under the proposed scheduling methods, we define the network average sum backlog as
\begin{align}\label{queue_ave}
\bar{U}&=\underset{T\rightarrow \infty}{\lim} \frac{1}{T}\sum^{T-1}_{t=0}\sum_{i=0}^{N}U_{i}{(t)}\nonumber\\
&=\sum_{i=0}^{N}\left\{\underset{T\rightarrow \infty}{\lim} \frac{1}{T}\sum^{T-1}_{t=0}U_{i}{(t)}\right\}=\sum_{i=0}^{N}\bar{U}_i
\end{align}
where $\bar{U}_i=\underset{T\rightarrow \infty}{\lim} \frac{1}{T}\sum^{T-1}_{t=0}U_{i}{(t)}$ denotes the time average backlog (the average number of packets) in the queue of node $i$. The backlog at the destination node $U_{N+1}{(t)}$ is always zero.

Note that if the average input rate at the source node $\bar{A}_0=\frac{1}{T}\sum_{t=0}^{T-1}A_0(t)$ exceeds the network capacity $\sfC_{\clL}$ ($\sfC_{\clD}$), the network would become unstable regardless of the underlying scheduling methods. However, within the network capacity range, a superior beam scheduling method should achieve a smaller average backlog \eqref{queue_ave} and therefore a smaller end-to-end delay (by Little's theorem). Here the end-to-end delay refers to the time taken for a packet to be transmitted across a network from the source node (node $0$) to the destination node (node $N+1$). The end-to-end delay comes from several sources including transmission delay, propagation delay, processing delay and queuing delay. We assume that the slot duration is long enough such that the afore mentioned transmission, propagation and processing time are included within each slot, and the slot duration remains constant regardless of the coding and scheduling policies. It follow that the most time consuming part is the queuing delay \cite{Park2010MSB}. By Little's theorem \cite{Little1961}, the average queuing delay time $\hat{W}$ that a packet spends in the network reads $\hat{W} = \bar{U}/\hat{A}$, where $\hat{A}$ is the long-term average effective arrival rate with $\hat{A}= \frac{1}{T}\sum^{T-1}_{t=0}A_0{(t)}=\bar{A}_0$.
In fact, the packet end-to-end delay $\hat{W}$ can also be decomposed into the average delays over each paths in the diamond network. More precisely, in the diamond network, each node $i\leq N$ has only one incident path with one queue of multiple packets. Denoted by $\bar{A}_i=\frac{1}{T}\sum^{T-1}_{t=0}A_i{(t)}$ as the long-term average arrival rate at node $i$, $\bar{W}_i$ as the average time that a packet spends in the queue of node $i$, based on Little's law we have $\bar{U}_i=\bar{A}_i\cdot \bar{W}_i$ with respect to (w.r.t.) the incident path of each node. Now we can write $\bar{W}$ as 
\begin{align}\label{pathpmf}
\hat{W} &= \frac{\bar{U}}{\hat{A}}=\sum_{i=0}^{N}\frac{\bar{A}_i}{\hat{A}}\bar{W}_i= \sum_{i=0}^{N}P_{i}\bar{W}_i,
\end{align}   
where we define $P_{i}=\frac{\bar{A}_i}{\hat{A}}$ as the activated time fraction (the probability mass function (PMF)) of the incident path at node $i$. With probability $1$ we have $P_0=1$ and $\sum_{i=1}^{N}P_{i}=1$. In the later simulation section, we will evaluate the network stability as well as the end-to-end delay performance w.r.t. different scheduling methods.

%

\section{Proposed Beam Scheduling Methods}\label{SchedulingMethod}
In this section, we will present three scheduling methods: the deterministic horizontal continuous edge coloring (HC-EC) scheduler, the adaptive back pressure (BP) scheduler and the adaptive low-delay back pressure (newBP) scheduler. The proposed schedulers are applicable for both of the line network and the diamond network, only with some minor modifications depending on the specific network characteristics.   
\subsection{The deterministic horizontal continuous edge coloring (HC-EC) scheduler}\label{HC-EC-alg}
\subsubsection{The HC-EC scheduler for the line network $\clL$} \label{HC_EC_line}
%
%
%
%
%
%
%
%
%
%
We consider a running example for the line network $\clL$, with $N=3$ and 
\begin{align}
l_{1}=8,\,\, l_{2}=8,\,\, l_{3}=12,\,\, l_{4} = 4,
\end{align}
where with a slight abuse of notation, we assume that the link capacity $l_{i}$ here are in the unit of packet per slot (packet/slot). Let $M$ be a common multiple of the link capacity $l_{i}$. We will construct an associate graph $\clG_{\clL}$ w.r.t. the line network $\clL$, where the set of nodes is the same as in $\clL$ and each link with capacity $l_{i}$ in $\clL$ is replaced by $n_{i}$ parallel edges, given by
\begin{align}
n_{i} = \frac{M}{l_{i}},\quad \forall i\in[N+1].
\end{align} 
Given the running example, we have 
\begin{align}
M=24,\,\, n_{1}=3,\,\, n_{2}=3,\,\, n_{3}=2,\,\, l_{4} = 6.
\end{align}
Since $\clG_{\clL}$ is a bipartite graph, namely, the vertices of $\clG_{\clL}$ can be divided into two disjoint and independent sets with set$\,1=\{0,2,4,...\}$ and set$\,2=\{1,3,5,...\}$ such that every edge connects a vertex in set$\,1$ to one in set$\,2$, we know that an optimal coloring can be performed with $\Delta_{\clL}$ colors, where $\Delta_{\clL}$ is the maximum node degree, given by
\begin{align}
\Delta_{\clL} = \max_{i\in[N]} \left\{n_{i}+n_{i+1}\right\}.
\end{align}
In the running example we have $\Delta_{\clL}=8$. Denoted by $\clC_{\clL}^{i}$ as the set of colors assigned to link $l_{i}$, $\Lambda_{\clL}\in\bC^{\Delta_{\clL}\times(N+1)}$ as the schedule matrix with each row corresponding to one network state $s$ w.r.t. link $l_{1},l_{2},\cdots,l_{N+1}$. Each state $s$ has an activated fraction of time $\omega_s=\frac{1}{\Delta_{\clL}}$. The HC-EC scheduler contains $\max\{n_i\}$ loops, and each loop starts from link $1$ until link $N+1$. Over each link $i$, the HC-EC scheduler extract one color with the minimum label from the currently available colors, and then assign this color to one edge at link $i$. For the running example, the coloring result is shown in TABLE$\,$\ref{tab_lin_ec} which contains $6$ loops and the schedule matrix $\Lambda_{\clL}$ w.r.t. the $\Delta_{\clL}$ network states reads
\begin{align}
\Lambda_{\clL}=\begin{bmatrix}
1 & 0 & 1 & 0  \\
0 & 1 & 0 & 1  \\
1 &  0 & 1& 0	\\
0 &  1 & 0& 1	\\
1 &  0 & 0& 1 	\\
0 & 1 & 0& 1 	\\
0 & 0 & 0& 1 	\\
0 & 0 & 0& 1 
\end{bmatrix},
\end{align}
where $\Lambda_{\clL, [c,i]}=1$, $c\in[\Delta_{\clL}]$, $i\in[N+1]$, indicates that link $i$ at $c$-th network state, i.e., $s=\Lambda_{\clL, [c,:]}$, is activated. Having obtained the network schedule matrix $\Lambda_{\clL}$, the network scheduler would be a simple deterministic repetition among the $\Delta_{\clL}$ states.

Note that, the EC method proposed in \cite{Yahya2017Line} tends to assign continuous interval of colors to each link, namely, the transmission of link $i$ can only begin when all the transmissions over link $i-1$ are finished, which results in large backlogs at node $i$. In contrast, the proposed HC-EC algorithm operates in multiple loops, for each loop, we color the link in a continuous horizontal order from link $1$ to link $N+1$ and always try to assign the minimum available color to each link. The intuition underlying the proposed HC-EC algorithm is that, any data packets entering into the network $\clL$ will be transmitted to the destination node as soon as possible. We will show later that the proposed HC-EC algorithm achieves much smaller end-to-end delay compared with the scheduler in \cite{Yahya2017Line}.
\begin{table}[t]
	\centering
	\caption{The HC-EC output for the line network $\clL$}
	\label{tab_lin_ec}
	\begin{tabular}{ |c|c|c|c|c| } 
		\hline
		Links $i$ & $i=1$ & $i=2$ & $i=3$ & $i=4$ \\
		\hline
		\multirow{6}{5em}{Assigned colors $\clC_{\clL}^{i}$} & 1 & 2  & 1  & 2\\ 
		& 3 & 4  & 3  & 4\\ 
		& 5 & 6  &    & 5\\ 
		&   &    &    & 6\\ 
		&   &    &    & 7\\ 
		&   &    &    & 8\\ 
		\hline
	\end{tabular}

\end{table}

\subsubsection{The HC-EC scheduler for the diamond network $\clD$}
%
%
%
%
%
%
%
%
%
%
By noticing that, each path in the diamond network can be regarded as a special case of $1$-relay line network $\clL$, in this section, we present the application of the HC-EC scheduler for the HD Gaussian diamond network $\clD$. 

We consider a running example for the diamond network $\clD$ with $N=4$ and the link capacities (packet/slot) are given by
\begin{align}
l_{1,1} = 3,\quad  & l_{1,2}=3\nonumber\\
l_{2,1} = 2,\quad  & l_{2,2}=3\nonumber\\
l_{3,1} = 3,\quad  & l_{3,2}=2\nonumber\\
l_{4,1} = 2,\quad  & l_{4,2}=2.
\end{align}
The activated fraction of time for each path can be obtained by \eqref{cap_diam}. For the running example, we have
\begin{align}
x_{1}=1,\,\, x_{2}=0.5,\,\, x_{3}=0.5,\,\, x_{4} = 0.
\end{align}
We assume that the total number of paths with non-zero activated time fraction is denoted by $P$. It follow that $P\leq \min\{3,N\}$ for the HD 1-2-1 diamond network $\clD$ as proved in \cite{Yahya2018Diam}. Without loss of generality, we will always assume that the first $P$ paths are activated with time fraction $x_{1}\geq x_{2}\geq \cdots \geq x_{P}$. In the running example, we have $P=3$. Let $M$ be a common multiple of the link capacity and path fraction time. Similar to the line network case, we will construct an associate graph $\clG_{\clD}$ such that the set of nodes coincide with the $P$ activated nodes in $\clD$ and each link with capacity $l_{p,j}$ in $\clD$ is replaced by $n_{p,j}$ parallel edges, given by
\begin{align}
n_{p,j} = M\cdot x_{p}\cdot \frac{l_{p,3-j}}{l_{p,1}+l_{p,2}}
\end{align} 
for all $p\in[P]$ and $j\in[2]$. For the running example, we have $M=10$ and 
\begin{align}
n_{1,1} = 5,\quad  & n_{1,2}=5\nonumber\\
n_{2,1} = 3,\quad  & n_{2,2}=2\nonumber\\
n_{3,1} = 2,\quad  & n_{3,2}=3
\end{align}
Again $\clG_{\clD}$ is a bipartite graph with two disjoint vertex sets, i.e., set$\,1=\{0,N+1\}$ and set$\,2=\{1,2,3,...N\}$, such that every edge connects a vertex in set$\,1$ to one in set$\,2$.
Now we can perform an optimal coloring with $\Delta_{\clD}$ colors, where $\Delta_{\clD}$ is the maximum node degree, given by
\begin{align}
\Delta_{\clD} = \max_{p\in[P]} \left\{\sum_{p'\in[P]}\!n_{p',1}, \sum_{p'\in[P]}\!n_{p',2}, \,\,  n_{p,1}\!+\!n_{p,2}\right\}.
\end{align}
In the running example, we have $\Delta_{\clD}=10$. Let $\clC_{\clD}^{p,j}$ denote the set of colors assigned to link $l_{p,j}$, $\Lambda_{\clD}\in\bC^{\Delta_{\clD}\times(2P)}$ denote the schedule with each row corresponding to one network state $s$ w.r.t. link $l_{1,1}, l_{1,2},l_{2,1},\cdots,l_{P,2}$ (from now on, we will ignore the non-active nodes and paths). Each state $s$ has an activated fraction of time $\omega_s=\frac{1}{\Delta_{\clD}}$. The HC-EC scheduler starts from path $1$ until path $P$. Over each path, the coloring procedure is exactly the same as for the line network in Section \ref{HC_EC_line}. For the running example, the coloring result is shown in TABLE$\,$\ref{tab_diam_ec} and the schedule matrix $\Lambda_{\clD}$ w.r.t. the $\Delta_{\clD}$ network states is given by
\begin{align}
\Lambda_{\clD}=\begin{bmatrix}
1 & 0 & 0 & 1 & 0 & 0  \\
0 & 1 & 1 & 0 & 0 & 0  \\
1 & 0 & 0 & 1 & 0 & 0  \\
0 & 1 & 1 & 0 & 0 & 0  \\
1 & 0 & 0 & 0 & 0 & 1  \\
0 & 1 & 1 & 0 & 0 & 0  \\
1 & 0 & 0 & 0 & 0 & 1  \\
0 & 1 & 0 & 0 & 1 & 0  \\
1 & 0 & 0 & 0 & 0 & 1  \\
0 & 1 & 0 & 0 & 1 & 0  
\end{bmatrix},
\end{align} 
\begin{table}[t]
	\centering
	\caption{The HC-EC output for the diamond network $\clD$}
	\label{tab_diam_ec}
	\begin{tabular}{ |c|c|c|c| } 
		\hline
		\multicolumn{2}{|c|}{Hops $j$} & $j=1$ & $j=2$ \\
		\hline
		\multirow{11}{5em}{Assigned colors $\clC_{\clD}^{p,j}$} & \multirow{5}{2em}{$\clC_{\clD}^{1,j}$} & 1  & 2 \\ 
		&  & 3  & 4  \\ 
		&  & 5 &  6  \\ 
		&   & 7   & 8   \\ 
		&   & 9   & 10   \\ 
		\cline{2-4}
		& \multirow{3}{2em}{$\clC_{\clD}^{2,j}$} & 2  & 1 \\ 
		&  &4  & 3  \\ 
		&  & 6 &    \\ 
		\cline{2-4}
		& \multirow{3}{2em}{$\clC_{\clD}^{3,j}$} & 8  & 5 \\ 
		&  & 10  & 7  \\ 
		&  &  &  9  \\ 
		\hline
	\end{tabular}
	
\end{table}
where $\Lambda_{\clD, [c,(p-1)\cdot 2+j]}=1$ indicates that the $j$-th link of path $p$ at $c$-th network state, i.e., $s=\Lambda_{\clD, [c,:]}$, is activated.

\textbf{Remark 1.} Note that for the HD 1-2-1 diamond network with $P\leq 3$ activated paths and time fraction $x_{1}\geq x_{2}\geq \cdots \geq x_{P}$, the aforementioned HC-EC scheduler is applicable only if the network $\clD$ satisfies one of the following conditions:

 {1) The activated number of paths $P\leq 2$;
 	
 {2) The activated number of paths $P=3$, with either $n_{3,1}\leq \max\{n_{1,2}-n_{2,1},0\}+\max\{n_{2,2}-n_{1,1},0\}$ or $n_{3,2}\leq \max\{n_{1,1}-n_{2,2},0\}+\max\{n_{2,1}-n_{1,2},0\}$.

In the case that the diamond network does not meet either of the above conditions, i.e., $P=3$ while the last path does not satisfies condition {2), the coloring procedure remains the same for the first $2$ paths. However, for the $3$-rd path, the scheduler should exchange $\min\{n_{3,1}, n_{3,2}\}$ colors between the remaining available colors for path $3$ and the union of already assigned colors for path $1$ and path $2$ at the same hop.

\subsection{The adaptive back pressure (BP) scheduler}
As illustrated in the above Section \ref{HC-EC-alg}, the HC-EC scheduler is rather simple, since once the schedule matrix $\Lambda_{\clL}$ ($\Lambda_{\clD}$) is obtained, the network activate states become deterministic. It follows that the scheduler just needs to periodically repeat the network states decided by $\Lambda_{\clL}$ ($\Lambda_{\clD}$). However, since the HC-EC scheduler is one-time predetermined by the network link capacities, the scheduler cannot handle seamlessly variations in the link capacities.
As an alternative approach, we can consider ``online'' dynamic scheduling policies that are guaranteed to achieve stability for all $\bar{A}_0<\clC_{\clL}$ (w.r.t. $\bar{A}_0<\clC_{\clD}$) without knowing explicitly the link capacities, and therefore can potentially adapt to (sufficiently slow) variations of the link capacities. In particular, we consider the well-known back pressure (BP) algorithm \cite{Neely2006backpressure} which is well understood to stabilize the network whenever the input rate lies within the capacity region of the network.

\subsubsection{The BP scheduler for the line network $\clL$}
Let $\Lambda_{\clL}(t)\in \bC^{N+1}$ denote the scheduling decision at slot $t$, with elements $\Lambda_{\clL}(t)_{[i]}=1$, $i\in[N+1]$, if link $i$ is scheduled$\,$/$\,$activated, otherwise $\Lambda_{\clL}(t)_{[i]}=0$. We define the differential backlog weight matrix $W(t)\in\bC^{N+1}$ with elements given by
\begin{align}
W(t)_{[i]} = \max\{U_{i-1}(t)-U_{i}(t), 0\}, \quad i\in[N+1]. 
\end{align}
Then choose the scheduling matrix $\Lambda_{\clL}(t)$ as the solution of the following binary integer programming (BIP)
\begin{align}\label{BIP_line}
\Lambda_{\clL}(t) &= \arg \max \sum_{i=1}^{N+1}W(t)_{[i]}\cdot \bar{r}_{i}(t)\cdot\Lambda_{\clL}(t)_{[i]}\nonumber\\
\quad &s.t. \quad \bar{r}_{i}(t) = \min\{U_{i-1}(t), l_{i}\}\nonumber\\
&\quad\quad\,\, \Lambda_{\clL}(t)_{[i]}\in\{0,1\}\nonumber\\
&\quad\quad\,\, \Lambda_{\clL}(t)_{[j]}+\Lambda_{\clL}(t)_{[j+1]}\leq 1, j\in[N],
\end{align}
where the first constraint indicates that each link rate at slot $t$ should not exceed the current backlog of the last departure node, the second constraint is the binary scheduling decision and the last constraint indicates the HD operating mode, i.e., the relay node cannot simultaneously transmit and receive. Consequently, the actual transmit rate for link $i$ is given by $r_{i}(t) = \Lambda_{\clL}(t)_{[i]}\cdot \bar{r}_{i}(t)$, and the slot-to-slot queuing evolution is given by \eqref{queue_dynamic} with $D_i(t)=r_{i+1}(t)$ and $A_{i}(t)=r_{i-1}(t)$. At node $0$ the number of arrival packets is the source input.

\subsubsection{The BP scheduler for the diamond network $\clD$}
For the diamond network, let $\Lambda_{\clD}(t)\in \bC^{N\times 2}$ denote the scheduling decision at slot $t$, the differential backlog weight matrix $W(t)\in\bC^{N\times2}$ reads
\begin{align}
W(t)_{[i,1]} &= \max\{U_{0}(t)-U_{i}(t), 0\},\\
W(t)_{[i,2]} &= U_{i}(t) - 0.
\end{align}
Then choose the scheduling matrix $\Lambda_{\clD}(t)$ as the solution of the following BIP optimization problem
\begin{align}\label{BIP_diam}
\Lambda_{\clD}(t) &= \arg \max \sum_{i=1}^{N}\sum_{j=1}^{2}W(t)_{[i,j]}\cdot \bar{r}_{i,j}(t)\cdot\Lambda_{\clD}(t)_{[i,j]}\nonumber\\
\quad &s.t. \quad \bar{r}_{i,1}(t) = \min\{U_{0}(t), l_{i,1}\}\nonumber\\
&\quad\quad\,\, \bar{r}_{i,2}(t) = \min\{U_{i}(t), l_{i,2}\}\nonumber\\
&\quad\quad\,\, \Lambda_{\clD}(t)_{[i,j]}\in\{0,1\}\nonumber\\
&\quad\quad\,\, \|\Lambda_{\clD}(t)_{[i,:]}\|_1\leq 1, i\in[N]\nonumber\\
&\quad\quad\,\, \|\Lambda_{\clD}(t)_{[:,j]}\|_1\leq 1, j\in[2],
\end{align}
where the first two constraints indicates that each link rate at slot $t$ should not exceed the current backlog of the last departure node, the third constraint is the binary scheduling decision, the fourth constraint indicates the HD operating mode, and the last constraint indicates that for each transmit (receive) the source (destination) node can only point its beam to one relay node in order to achieve the full beamforming gain \cite{sxs2019OSPS}. Consequently, the actual transmit rate for link $(i,j)$ is given by $r_{i,j}(t) = \Lambda_{\clD}(t)_{[i,j]}\cdot \bar{r}_{i,j}(t)$, and the queuing evolution is given by \eqref{queue_dynamic} with $D_i(t)=r_{i,2}(t)$ and $A_{i}(t)=r_{i,1}(t)$ for $i\in [N]$. At node $0$ the number of arrival packets is the source input while $D_0(t)=\sum_{i=1}^{N}r_{i,1}(t)$.

\textbf{Remark 2.} According to Edmonds' matching polytope theorem \cite{Edmonds1965}, one can easily prove that the BIP problem in \eqref{BIP_line} (w.r.t. \eqref{BIP_diam}) are identical to the {\em Matching polytope}. Namely we can replace the binary constraint $\Lambda_{\clL}(t)_{[i]}\in\{0,1\}$ in \eqref{BIP_line} (w.r.t. $\Lambda_{\clD}(t)_{[i,j]}\in\{0,1\}$ in \eqref{BIP_diam}) with the linear constraint $\Lambda_{\clL}(t)_{[i]}\in[0,1]$ (w.r.t. $\Lambda_{\clD}(t)_{[i,j]}\in[0,1]$), such that the BIP problem simply boils down to a linear programming with all the extreme points are binary integral.

\subsection{The adaptive low-delay back pressure (newBP) scheduler}
Several variants of BP have appeared in the recent literature \cite{Shroff2017, Neely2018newBP} with the goal of improving the delay properties of the basic BP scheme. In this section, we present the newBP beam scheduler, which is based on the state-of-the-art new BP algorithm proposed in \cite{Shroff2017}. In this new BP algorithm, the ``back pressure'' quantities are with respect to virtual queue lengths aiming at reducing the implementation complexity for large networks. Moreover, the new BP algorithm involves a quadratic term for a further reduction of the end-to-end delay.

\subsubsection{The newBP scheduler for the line network $\clL$}
Choose parameters $\rho>0$, $\tau>0$ and $\beta_{j}>0$, with $j\in [N+1]$. Let $U_{i}{(t)}$ and $V_{i}{(t)}$ denote the physical and virtual queues (backlog), respectively, which are empty at the initial state $U_{i}{(0)}=V_{i}{(0)}=V_{i}{(-1)}=0$. At the beginning of each slot $t$, calculate the new weight
\begin{align}
	z_i(t)=(1+\frac{1}{\tau})V_i(t-1)-\frac{1}{\tau}V_i(t-2),
\end{align}
where for the destination node, we have $z_{N+1}(t)\equiv 0$. Denoted by $\tilde{W}(t)\in\bC^{N+1}$ as the new differential backlog weight matrix with elements given by 
\begin{align}
	\tilde{W}(t)_{[i]} = z_{i-1}(t)-z_i(t).
\end{align}
Then choose the scheduling matrix $\Lambda_{\clL}(t)$ as the solution of the following BIP problem
\begin{align}
\Lambda_{\clL}(t)&= \arg \max \sum_{i=1}^{N+1}\tilde{W}(t)_{[i]} r_i(t)-\frac{\rho\beta_{i}}{2}\left[r_i(t)-r_i(t-1)\right]^2\nonumber\\
\quad &s.t. \quad r_i(t) = \bar{r}_{i}(t)\cdot\Lambda_{\clL}(t)_{[i]}\nonumber\\
&\quad\quad\,\, \bar{r}_{i}(t) = \min\{U_{i-1}(t), l_{i}\}\nonumber\\
&\quad\quad\,\, \Lambda_{\clL}(t)_{[i]}\in\{0,1\}\nonumber\\
&\quad\quad\,\, \Lambda_{\clL}(t)_{[j]}+\Lambda_{\clL}(t)_{[j+1]}\leq 1, j\in[N],
\end{align}
where the constraints indicates the same meaning as in \eqref{BIP_line}. 

Finally at the end of each time slot $t$, update the virtual queue length $V_i(t)$ by
\begin{align}
	V_i(t) = V_i(t-1)-\rho\tau D_i(t)+\rho\tau A_i(t), 
\end{align}
where $D_i(t)=r_{i+1}(t)$, $A_{i}(t)=r_{i-1}(t)$ and at node $0$ the number of arrival packets is the source input. The evolution of the physical queue $U_{i}{(t)}$ is given by \eqref{queue_dynamic}.

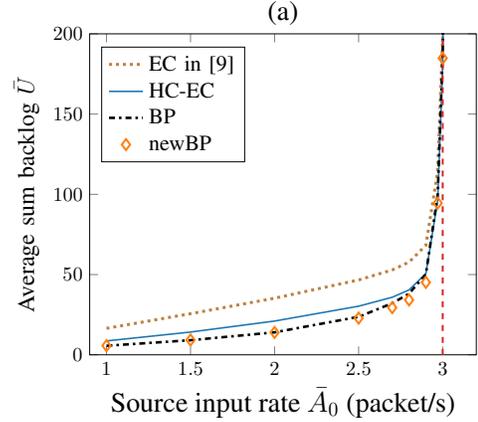
\begin{figure}[t]
	\centering
	\scalebox{0.75}{
\begin{tikzpicture}

\definecolor{color0}{rgb}{0.12156862745098,0.466666666666667,0.705882352941177}
\definecolor{color1}{rgb}{1,0.498039215686275,0.0549019607843137}
\definecolor{color2}{rgb}{0.172549019607843,0.627450980392157,0.172549019607843}
\definecolor{color3}{rgb}{0.83921568627451,0.152941176470588,0.156862745098039}
\definecolor{color4}{rgb}{0.580392156862745,0.403921568627451,0.741176470588235}

\begin{axis}[%
xmin=0.9,
xmax=3.2,
xlabel={\Large Source input rate $\bar{A}_0$ (packet/s)},
ymin=0.0,
ymax=200,
ytick={0,50,100,150,200},
yminorticks=true,
ylabel={\large Average sum backlog $\bar{U}$},
title style={at={(0.5,0.97)}},
title = {\Large (a)},
legend style={at={(0.03,0.60)},nodes={scale=1.1, transform shape}, anchor=south west, legend cell align=left, align=left, draw=black}
]
\addplot [line width=1.5pt, brown, dotted]
table {%
1 16.4862570028011
1.5 25.6136329531813
2 35.2587159863946
2.5 46.6463522909164
2.7 52.8963272809124
2.8 57.7586534613846
2.9 68.2439475790316
2.97 114.1237494998
3 198.266218987595
3.2 2561.28290691277
};
\addlegendentry{EC in \cite{Yahya2017Line}}
\addplot [line width=0.8pt, color0]
table {%
1 8.63621698679472
1.5 14.1753701480592
2 21.0244410264106
2.5 30.2953931572629
2.7 35.8534351240496
2.8 40.4433648459384
2.9 50.6132265406163
2.97 95.9943633703481
3 180.151013530412
3.2 2543.03710234094
};
\addlegendentry{HC-EC}
\addplot [line width=1.3pt, black, dashdotted]
table {%
1 5.62677571028411
1.5 9.10190951380552
2 13.9970988395358
2.5 23.6711309523809
2.7 32.1175157563025
2.8 38.124724889956
2.9 50.2708270808323
2.97 99.1284451280512
3 186.874781162465
3.2 2668.74303471389
};
\addlegendentry{BP}
\addplot [line width=1.0pt, color1, mark=diamond, mark size=2.8, mark options={solid,fill opacity=0}, only marks]
table {%
1 5.68829406762705
1.5 9.2390893857543
2 13.9463660464186
2.5 22.8611819727891
2.7 29.390524959984
2.8 34.1800032513005
2.9 45.1474839935974
2.97 94.5200486444578
3 184.834887079832
3.2 2754.78128751501
};
\addlegendentry{newBP}
\addplot [line width=1.0pt, color3, dashed, forget plot]
table {%
3 0
3 200
};
\end{axis}

\end{tikzpicture}}
	\caption{The time averaged backlog $\bar{U}$ with respect to different source input rates $\bar{A}_0$ for the line network $\clL$. For the newBP scheduler, we set $\rho=\tau=1$, $\beta_1=0.3$, $\beta_2=0.2$, $\beta_3=0.1$, $\beta_4=0$. }
	\label{line_ave_queue}
\end{figure}
\begin{figure}[t]
	\centering
	\scalebox{0.75}{
\begin{tikzpicture}

\definecolor{color0}{rgb}{0.12156862745098,0.466666666666667,0.705882352941177}
\definecolor{color1}{rgb}{1,0.498039215686275,0.0549019607843137}
\definecolor{color2}{rgb}{0.172549019607843,0.627450980392157,0.172549019607843}
\definecolor{color3}{rgb}{0.83921568627451,0.152941176470588,0.156862745098039}
\definecolor{color4}{rgb}{0.580392156862745,0.403921568627451,0.741176470588235}

\begin{axis}[%
xmin=0.9,
xmax=2.9,
xlabel={\Large Source input rate $\bar{A}_0$ (packet/s)},
ymin=0.0,
ymax=175,
ytick={0,50,100,150,200},
yminorticks=true,
ylabel={\large Average sum backlog $\bar{U}$},
title style={at={(0.5,0.97)}},
title = {\Large (b)},
legend style={at={(0.03,0.69)},nodes={scale=1.1, transform shape}, anchor=south west, legend cell align=left, align=left, draw=black}
]
\addplot [line width=0.8pt, color0]
table {%
1 3.45732629238466
1.75 6.57471928849361
2.25 9.90004446914953
2.55 17.5302612562535
2.6 21.7389605336298
2.65 29.5538465814341
2.675 56.6539577543079
2.7 174.777759866593
2.8 889.971484157866
};
\addlegendentry{HC-EC}
\addplot [line width=1.3pt, black, dashdotted]
table {%
1 2.12605892162312
1.75 4.68646470261256
2.25 9.37898276820456
2.55 22.2818065591996
2.6 29.4165591995553
2.65 42.3734797109505
2.675 86.1276209005003
2.7 267.909749861034
2.8 1332.93752084491
};
\addlegendentry{BP}
\addplot [line width=1.0pt, color1, mark=diamond, mark size=2.8, mark options={solid,fill opacity=0}, only marks]
table {%
1 2.06956086714842
1.75 4.20206225680934
2.25 8.31430239021679
2.55 20.7748249027237
2.6 27.8569760978321
2.65 39.9679599777654
2.675 81.7696053362979
2.7 259.070355753196
2.8 1324.65265147304
};
\addlegendentry{newBP}
\addplot [line width=1.0pt, color3, dashed, forget plot]
table {%
2.7 0
2.7 200
};
\end{axis}

\end{tikzpicture}}
	\caption{The time averaged backlog $\bar{U}$ with respect to different source input rates $\bar{A}_0$ for the diamond network $\clD$. For the newBP scheduler, we set $\rho=\tau=1$, $\beta_{1,1}=0$, $\beta_{2,1}=1$, $\beta_{3,1}=2$, $\beta_{4,1}=16$, and $\beta_{i,j}=0$ for $j=2$.}
	\label{diam_ave_queue}
\end{figure}
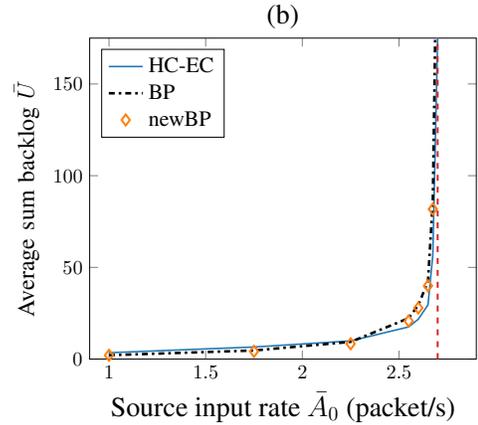
\begin{figure*}[t]
	\centering
	\scalebox{0.65}{\input{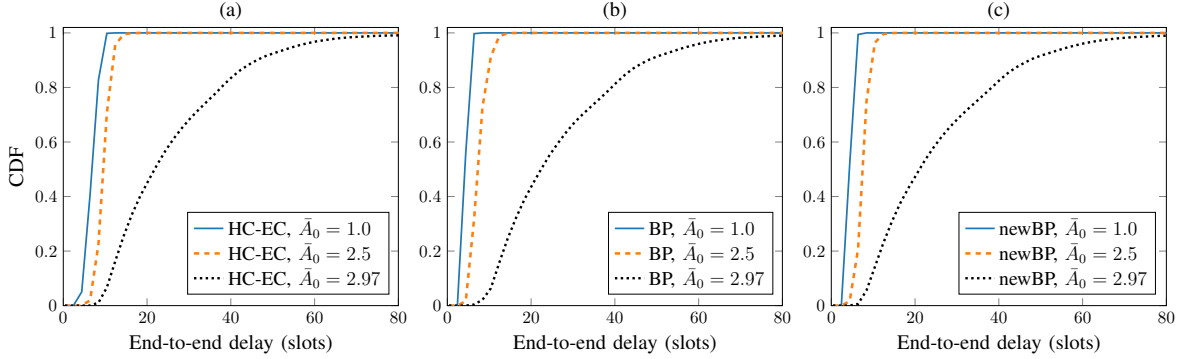}}
	\caption{The distribution of the packet end-to-end delay in terms of different source input rates $\bar{A}_0$ for the line network $\clL$: (a) The HC-EC scheduler, (b) The BP scheduler, (c) The newBP scheduler.}
	\label{line_n2nDelay}
\end{figure*}
\subsubsection{The newBP scheduler for the diamond network $\clD$}
For the diamond network $\clD$, choose parameters $\rho>0$, $\tau>0$ and $\beta_{i,j}>0$, with $i\in [N]$ and $j\in [2]$. Denoted by $U_{i}{(t)}$ and $V_{i}{(t)}$ as the physical and virtual queues (backlog), respectively, with $i=0$ indicating the source node and $i=N+1$ indicating the destination node, otherwise indicating the relay nodes. Let $U_{i}{(t)}$ and $V_{i}{(t)}$ be empty at the initial state $U_{i}{(0)}=V_{i}{(0)}=V_{i}{(-1)}=0$. At the beginning of each slot $t$, calculate the new weights
\begin{align}
z_i(t)=(1+\frac{1}{\tau})V_i(t-1)-\frac{1}{\tau}V_i(t-2),
\end{align}
where for the destination node, we have $z_{N+1}(t)\equiv 0$. Denoted by $\tilde{W}(t)\in\bC^{N\times 2}$ as the new differential backlog weight matrix with elements given by 
\begin{align}
\tilde{W}(t)_{[i,1]} &= z_{0}(t)-z_i(t),\\
\tilde{W}(t)_{[i,2]} &= z_i(t)
\end{align}
for the first and the second hops, respectively. Then choose the scheduling matrix $\Lambda_{\clD}(t)$ as the solution of the following BIP optimization problem
\begin{align}
\Lambda_{\clD}(t) = & \arg \max \!\!\!\!\!\!\!\!\sum_{i\in[N],j\in[2]}\!\!\!\!\!\!\!\!\tilde{W}(t)_{[i,j]}r_{i,j}(t)-\frac{\rho\beta_{i,j}}{2}\!\!\left[r_{i,j}(t)-r_{i,j}(t\!-\!1)\right]^2\nonumber\\
\quad &s.t. \quad r_{i,j}(t) = \bar{r}_{i,j}(t)\cdot\Lambda_{\clD}(t)_{[i,j]}\nonumber\\
&\quad\quad\,\, \bar{r}_{i,1}(t) = \min\{U_{0}(t), l_{i,1}\}\nonumber\\
&\quad\quad\,\, \bar{r}_{i,2}(t) = \min\{U_{i}(t), l_{i,2}\}\nonumber\\
&\quad\quad\,\, \Lambda_{\clD}(t)_{[i,j]}\in\{0,1\}\nonumber\\
&\quad\quad\,\, \|\Lambda_{\clD}(t)_{[i,:]}\|_1\leq 1, i\in[N]\nonumber\\
&\quad\quad\,\, \|\Lambda_{\clD}(t)_{[:,j]}\|_1\leq 1, j\in[2],
\end{align}
where the first constraint indicates the actual transmit rate for link $(i,j)$ at slot $t$, the second and the third constraints indicates that each link rate at slot $t$ should not exceed the current backlog of the last departure node, the fourth constraint is the binary scheduling decision, the fifth constraint indicates the HD operating mode and the last constraint indicates that for each transmit (receive) the source (destination) node can only point its beam to one relay node in order to achieve the full beamforming gain \cite{sxs2019OSPS}.

Finally at the end of each time slot $t$, update the virtual queue length $V_i(t)$ by
\begin{align}
V_i(t) = V_i(t-1)-\rho\tau D_i(t)+\rho\tau A_i(t),
\end{align}
where $D_i(t)=r_{i,2}(t)$ and $A_{i}(t)=r_{i,1}(t)$ for $i\in [N]$. At node $0$ the number of arrival packets is the source input while $D_0(t)=\sum_{i=1}^{N}r_{i,1}(t)$. The evolution of the physical queue $U_{i}{(t)}$ is given by \eqref{queue_dynamic}.

\begin{figure*}[h]
	\centering
	\scalebox{0.65}{\input{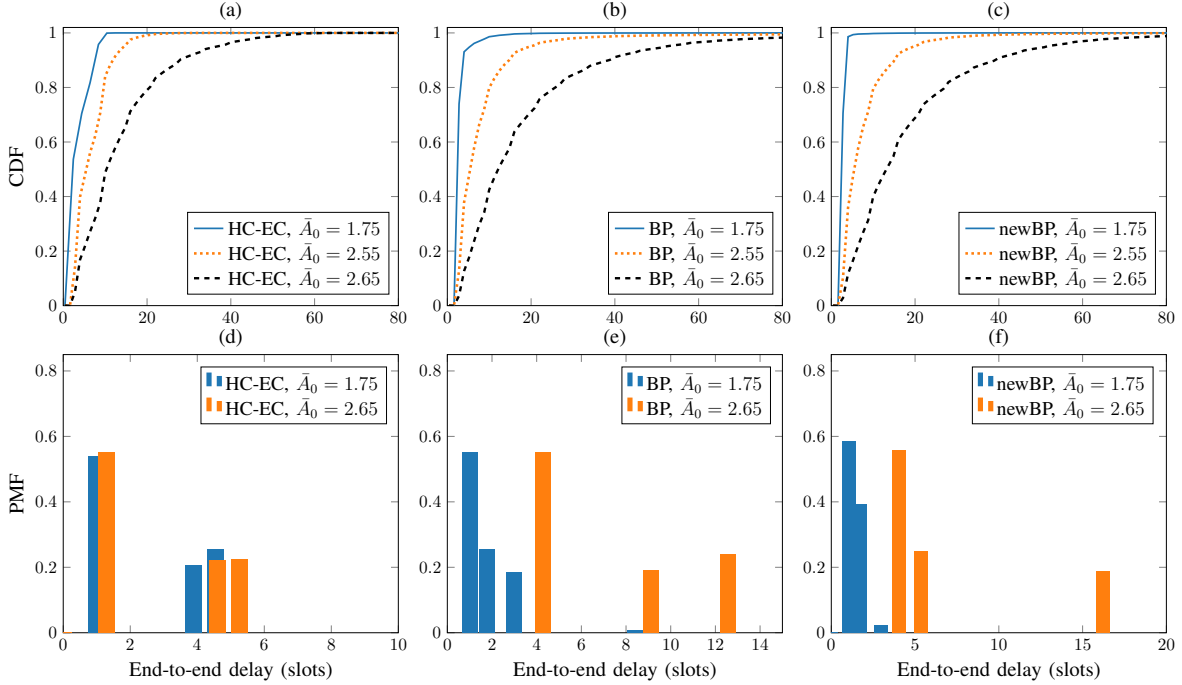}}
	\caption{The distribution of the packet end-to-end delay in terms of different source input rates $\bar{A}_0$ and in terms of different relay paths for the diamond network $\clD$: (a)(d) The HC-EC scheduler, (b)(e) The BP scheduler, (c)(f) The newBP scheduler.}
	\label{diam_n2nDelay}
\end{figure*}
\begin{figure*}[t]
	\centering
	\scalebox{0.75}{
\begin{tikzpicture}

\definecolor{color0}{rgb}{0.12156862745098,0.466666666666667,0.705882352941177}
\definecolor{color1}{rgb}{1,0.498039215686275,0.0549019607843137}
\definecolor{color2}{rgb}{0.172549019607843,0.627450980392157,0.172549019607843}

\begin{groupplot}[group style={group size=2 by 1}]
\nextgroupplot[
xmin=7,
xmax=30,
xlabel={\Large End-to-end delay (slots)},
ymin=0.8,
ymax=1.004,
yminorticks=true,
ylabel={\large  CDF},
title style={at={(0.5,0.97)}},
title = {\Large (a)},
legend style={at={(0.40,0.04)},nodes={scale=1.1, transform shape}, anchor=south west, legend cell align=left, align=left, draw=black}
]
\addplot [line width=1.4pt, black, dotted]
table {%
0.4 0
1.6 0
2.8 0
4 7.60901710920956e-05
5.2 0.0022792268834669
6.4 0.019581872380732
7.6 0.0820696281081413
8.8 0.225749282883623
10 0.705627361818802
11.2 0.875613005500055
12.4 0.95583046753836
13.6 0.985325495328628
14.8 0.994950307812045
16 0.999479439287389
17.2 0.99981975987824
18.4 0.999937967595365
19.6 0.999990003198976
20.8 1
22 1
23.2 1
24.4 1
25.6 1
26.8 1
28 1
29.2 1
30.4 1
31.6 1
32.8 1
34 1
35.2 1
36.4 1
37.6 1
38.8 1
40 1
41.2 1
42.4 1
43.6 1
44.8 1
46 1
47.2 1
48.4 1
49.6 1
50.8 1
52 1
53.2 1
54.4 1
55.6 1
56.8 1
58 1
59.2 1
60.4 1
61.6 1
62.8 1
64 1
65.2 1
66.4 1
67.6 1
68.8 1
70 1
71.2 1
72.4 1
73.6 1
74.8 1
76 1
77.2 1
78.4 1
79.6 1
80.8 1
82 1
83.2 1
84.4 1
85.6 1
86.8 1
88 1
89.2 1
90.4 1
91.6 1
92.8 1
94 1
95.2 1
96.4 1
97.6 1
98.8 1
100 1
101.2 1
102.4 1
103.6 1
104.8 1
106 1
107.2 1
108.4 1
109.6 1
110.8 1
112 1
113.2 1
114.4 1
115.6 1
116.8 1
118 1
119.2 1
120.4 1
121.6 1
122.8 1
124 1
125.2 1
126.4 1
127.6 1
128.8 1
130 1
131.2 1
132.4 1
133.6 1
134.8 1
136 1
137.2 1
138.4 1
139.6 1
140.8 1
142 1
143.2 1
144.4 1
145.6 1
146.8 1
148 1
149.2 1
150.4 1
151.6 1
152.8 1
154 1
155.2 1
156.4 1
157.6 1
158.8 1
160 1
};
\addlegendentry{HC-EC, $\bar{A}_0=2.5$}
\addplot [line width=1.5pt, color1, dashed]
table {%
0.4 0
1.6 0
2.8 0
4 0.0164022247064803
5.2 0.11394702955391
6.4 0.318483405873474
7.6 0.547623432400373
8.8 0.728914764519897
10 0.913786140175774
11.2 0.962737312696384
12.4 0.988176092066004
13.6 0.996304323322996
14.8 0.998941542213959
16 0.999947913954365
17.2 1
18.4 1
19.6 1
20.8 1
22 1
23.2 1
24.4 1
25.6 1
26.8 1
28 1
29.2 1
30.4 1
31.6 1
32.8 1
34 1
35.2 1
36.4 1
37.6 1
38.8 1
40 1
41.2 1
42.4 1
43.6 1
44.8 1
46 1
47.2 1
48.4 1
49.6 1
50.8 1
52 1
53.2 1
54.4 1
55.6 1
56.8 1
58 1
59.2 1
60.4 1
61.6 1
62.8 1
64 1
65.2 1
66.4 1
67.6 1
68.8 1
70 1
71.2 1
72.4 1
73.6 1
74.8 1
76 1
77.2 1
78.4 1
79.6 1
80.8 1
82 1
83.2 1
84.4 1
85.6 1
86.8 1
88 1
89.2 1
90.4 1
91.6 1
92.8 1
94 1
95.2 1
96.4 1
97.6 1
98.8 1
100 1
101.2 1
102.4 1
103.6 1
104.8 1
106 1
107.2 1
108.4 1
109.6 1
110.8 1
112 1
113.2 1
114.4 1
115.6 1
116.8 1
118 1
119.2 1
120.4 1
121.6 1
122.8 1
124 1
125.2 1
126.4 1
127.6 1
128.8 1
130 1
131.2 1
132.4 1
133.6 1
134.8 1
136 1
137.2 1
138.4 1
139.6 1
140.8 1
142 1
143.2 1
144.4 1
145.6 1
146.8 1
148 1
149.2 1
150.4 1
151.6 1
152.8 1
154 1
155.2 1
156.4 1
157.6 1
158.8 1
160 1
};
\addlegendentry{BP, $\,\,\bar{A}_0=2.5$}
\addplot [line width=1.0pt, color0]
table {%
0.4 0
1.6 0
2.8 0
4 0.0164534081051143
5.2 0.0640454756986365
6.4 0.20884253347251
7.6 0.481313867274014
8.8 0.749327445181996
10 0.964313013333043
11.2 0.987689060657339
12.4 0.995962918226609
13.6 0.998641730254515
14.8 0.999551452797197
16 0.999967992941202
17.2 0.999996001759226
18.4 1
19.6 1
20.8 1
22 1
23.2 1
24.4 1
25.6 1
26.8 1
28 1
29.2 1
30.4 1
31.6 1
32.8 1
34 1
35.2 1
36.4 1
37.6 1
38.8 1
40 1
41.2 1
42.4 1
43.6 1
44.8 1
46 1
47.2 1
48.4 1
49.6 1
50.8 1
52 1
53.2 1
54.4 1
55.6 1
56.8 1
58 1
59.2 1
60.4 1
61.6 1
62.8 1
64 1
65.2 1
66.4 1
67.6 1
68.8 1
70 1
71.2 1
72.4 1
73.6 1
74.8 1
76 1
77.2 1
78.4 1
79.6 1
80.8 1
82 1
83.2 1
84.4 1
85.6 1
86.8 1
88 1
89.2 1
90.4 1
91.6 1
92.8 1
94 1
95.2 1
96.4 1
97.6 1
98.8 1
100 1
101.2 1
102.4 1
103.6 1
104.8 1
106 1
107.2 1
108.4 1
109.6 1
110.8 1
112 1
113.2 1
114.4 1
115.6 1
116.8 1
118 1
119.2 1
120.4 1
121.6 1
122.8 1
124 1
125.2 1
126.4 1
127.6 1
128.8 1
130 1
131.2 1
132.4 1
133.6 1
134.8 1
136 1
137.2 1
138.4 1
139.6 1
140.8 1
142 1
143.2 1
144.4 1
145.6 1
146.8 1
148 1
149.2 1
150.4 1
151.6 1
152.8 1
154 1
155.2 1
156.4 1
157.6 1
158.8 1
160 1
};
\addlegendentry{newBP, $\bar{A}_0=2.5$}

\nextgroupplot[
xmin=35,
xmax=120,
xlabel={\Large End-to-end delay (slots)},
ymin=0.8,
ymax=1.004,
yminorticks=true,
title style={at={(0.5,0.97)}},
title = {\Large (b)},
legend style={at={(0.37,0.04)},nodes={scale=1.1, transform shape}, anchor=south west, legend cell align=left, align=left, draw=black}
]
\addplot [line width=1.4pt, black, dotted]
table {%
0.4 0
4.4 0
8.4 0.0107612760851712
12.4 0.1625876391875
16.4 0.334677787607315
20.4 0.460536848407106
24.4 0.568188698740604
28.4 0.652637710297235
32.4 0.719638431392699
36.4 0.779213897093356
40.4 0.840086724916962
44.4 0.884544673106705
48.4 0.915707549779258
52.4 0.936177232754145
56.4 0.955059014803311
60.4 0.968907643677379
64.4 0.977897580775294
68.4 0.983857335454395
72.4 0.987229205570006
76.4 0.989905725614537
80.4 0.991093098989052
84.4 0.99321769994475
88.4 0.997421687957274
92.4 0.999273384787959
96.4 0.999961492742219
100.4 1
104.4 1
108.4 1
112.4 1
116.4 1
120.4 1
124.4 1
128.4 1
132.4 1
136.4 1
140.4 1
144.4 1
148.4 1
152.4 1
156.4 1
};
\addlegendentry{HC-EC, $\bar{A}_0=2.97$}
\addplot [line width=1.5pt, color1, dashed]
table {%
0.4 0
4.4 0.000106202930336923
8.4 0.0214627065855569
12.4 0.157329322035324
16.4 0.321796910041024
20.4 0.446387904187201
24.4 0.551546614075231
28.4 0.636224354156641
32.4 0.702913870045128
36.4 0.758950127385744
40.4 0.818147809539525
44.4 0.86979527032483
48.4 0.903432223545186
52.4 0.925394497092178
56.4 0.94541785302047
60.4 0.961255726058387
64.4 0.9727105598555
68.4 0.980241860945098
72.4 0.984928288882687
76.4 0.987957291716903
80.4 0.989967275238459
84.4 0.99105232314776
88.4 0.992899120971117
92.4 0.997049811636668
96.4 0.999059020510674
100.4 0.999892842193386
104.4 1
108.4 1
112.4 1
116.4 1
120.4 1
124.4 1
128.4 1
132.4 1
136.4 1
140.4 1
144.4 1
148.4 1
152.4 1
156.4 1
};
\addlegendentry{BP, $\,\,\bar{A}_0=2.97$}
\addplot [line width=1.0pt, color0]
table {%
0.4 0
4.4 0.000222543142730237
8.4 0.0556489229261401
12.4 0.228592194435104
16.4 0.373785399303775
20.4 0.482509069361278
24.4 0.579519468333616
28.4 0.656427264995493
32.4 0.717805559021877
36.4 0.772846663112595
40.4 0.829999286906265
44.4 0.875925628164003
48.4 0.907688052472939
52.4 0.927983283971116
56.4 0.946864004496241
60.4 0.96194093470511
64.4 0.972505987080724
68.4 0.979969710004583
72.4 0.984699278061144
76.4 0.987642755901562
80.4 0.989914470721187
84.4 0.990962665327306
88.4 0.992514649254981
92.4 0.996400468776159
96.4 0.998955298844801
100.4 0.999800770132262
104.4 1
108.4 1
112.4 1
116.4 1
120.4 1
124.4 1
128.4 1
132.4 1
136.4 1
140.4 1
144.4 1
148.4 1
152.4 1
156.4 1
};
\addlegendentry{newBP, $\bar{A}_0=2.97$}
\end{groupplot}

\end{tikzpicture}}
	\caption{The comparison of the proposed three beam schedulers in terms of the distribution of the packet end-to-end delay for the line network $\clL$.}
	\label{line_delayCompare}
\end{figure*}
\begin{figure*}[t]
	\centering
	\scalebox{0.75}{\input{3_diam_compare.tex}}
	\caption{The comparison of the proposed three beam schedulers in terms of the distribution of the packet end-to-end delay for the diamond network $\clD$}
	\label{diam_delayCompare}
\end{figure*}
\section{Numerical Results}\label{simulations}
In this section, we investigate the numerical performance of the proposed HC-EC, BP and newBP beam scheduling methods. We consider the same running example as in Section \ref{SchedulingMethod} with $N=4$, $l_1=8$, $l_2=8$, $l_3=12$, $l_4=4$ for the line network $\clL$, and $N=4$, $l_{1,1}=3$, $l_{1,2}=3$, $l_{2,1}=2$, $l_{2,2}=3$, $l_{3,1}=3$, $l_{3,2}=2$, $l_{4,1}=2$, $l_{4,2}=2$ for the diamond network $\clD$, respectively. The information theoretic capacities for the line network $\clL$ and the diamond network $\clD$ are given by \eqref{cap_line_simp} and \eqref{cap_diam} with $\sfC_{\clL}=3$ and $\sfC_{\clD}=2.7$. We assume that the number of packets arriving at the source node at each time slot obey the Poisson distribution with mean $\bar{A}_0$.

\subsection{The evaluation of network stability}
We first investigate the network stability. For the line network $\clL$ as shown in \figref{line_ave_queue}, we increase the source input rates $\bar{A}_0$ and compare the long-term time average backlog (queue lengths) w.r.t. different beam schedulers. As we can see, within the network capacity range $[0,\clC_{\clL}]$, all the physical queue lengths with respect to the proposed three schedulers (i.e., the HC-EC scheduler, the BP scheduler and the newBP scheduler) are finite, which indicates the guarantee for the network stability. In particular, the average queue backlog of the proposed schedulers are significantly smaller than that achieved by the original EC scheduler in \cite{Yahya2017Line} for rates $\bar{A}_0$ not too close to $\clC_{\clL}$.

To the best of our knowledge, there are no reference baseline schedulers for the diamond or similar mmWave relay networks in the literature. Therefore, we can only present simulations comparing our proposed approaches. As shown in \figref{diam_ave_queue}, within the network capacity range $[0,\clC_{\clD}]$, 
the average backlog achieved by the proposed schedulers are very similar, concluding that for the diamond 1-2-1 network these methods are quite equivalent and they all achieve stability for all $\bar{A}_0<\clC_{\clD}$, in accordance with the theory.

\subsection{The evaluation of end-to-end delay}
In this section, we investigate the packet end-to-end delay w.r.t. the proposed three schedulers. Here the end-to-end delay indicates how long the packets are delayed in the queues during the transmission from source node to the destination node. For the line network $\clL$ as shown in \figref{line_n2nDelay}, the cumulative density function (CDF) of the packet delay indicates the probability that packet end-to-end delay is smaller than the specified delay. As we can see, for any of the proposed three schedulers, by increasing the source input rates from $\bar{A}_0=1$ to $\bar{A}_0=2.5$ and $\bar{A}_0=2.97$, the CDF curve translates to the right. Namely, by increasing the source input rate, the packets experience longer delays. For the diamond network $\clD$, we obtain the similar results as shown in \figref{diam_n2nDelay}$\,$(a)-(c). Particularly, we also evaluate the delay performance with respect to different relay paths as discussed in \eqref{pathpmf}. As we can see from \figref{diam_n2nDelay}$\,$(d)-(f), each relay path has its own delay distribution. By increasing the source input rate, the delay distribution translates to the right side which indicates a longer packet delay.

\figref{line_delayCompare} compares the delay performance of the proposed three beam schedulers w.r.t. the line network $\clL$. As we can see from \figref{line_delayCompare}$\,$(a), with a low source input rate $\bar{A}_0=2.5$, the packets with the newBP scheduler experience the smallest delays, followed by the BP scheduler and the HC-EC scheduler. When the source input rate is close to the network capacity $\bar{A}_0=2.97$, the schedulers perform very similar with the HC-EC scheduler slightly superior than the other two. The proposed schedulers w.r.t. the diamond network achieve similar performance. As shown in \figref{diam_delayCompare}, the newBP scheduler outperforms the other two when the source input is very low. By increasing the source input rate, the HC-EC scheduler performance better than the other two. 

Note that, each of the proposed three schedulers has its own advantages in terms of different implementation regardings. From the computation complexity point of view, the HC-EC scheduler is preferred since it is one-time computation and then repetition, which results in less adaptivity to channel changes. The computation complexity of the adaptive BP scheduler is slightly less than the adaptive newBP scheduler since the former is identical to the {\em Matching polytope} and can be solved by a simple linear programming. Regarding the packet end-to-end delay, in low source input rate range, the newBP scheduler performance better than the other two, while near to the network capacity, the HC-EC scheduler outperforms the others.

\section{Conclusion}\label{conclusion}
In this paper, we focused on two typical relay models, i.e., the line network and the diamond network, which provides the basic topology for any further more complex mmWave relay networks. We proposed three beam schedulers to approach the network information theoretic capacity, i.e., the deterministic horizontal continuous edge coloring (HC-EC) scheduler, the adaptive back pressure (BP) scheduler and the adaptive low-delay new back pressure (newBP) scheduler. Within the network capacity range, all the schedulers can guarantee the network stability and achieve very low packet delay. Particularly, we have shown that each scheduler has its own advantages in terms of different implementation regardings.

\section*{Acknowledgments}
X. Song is sponsored by the China Scholarship Council (201604910530). The authors would like to thank Yahya H. Ezzeldin, Prof. Christina Fragouli and Prof. Michael J. Neely, who provide valuable discussions.

\balance
{\footnotesize
	\bibliographystyle{IEEEtran}
	\bibliography{references}
}

\end{document}